\documentclass[aps,pra,reprint,showpacs,longbibliography,superscriptaddress]{revtex4-1}
\usepackage{bm}
\usepackage{graphicx}
\usepackage{epstopdf}
\usepackage{latexsym}
\usepackage{subfigure}
\usepackage{color}
\usepackage{hyperref}
\usepackage{amssymb}
\usepackage{amsmath}
\usepackage{amsbsy,bm}
\renewcommand{\vec}[1]{{\bm{#1}}}
\usepackage{fancyhdr}

\renewcommand{\vec}[1]{{\bm{#1}}}

\begin{document}

\title{Microwave dynamics of pure and doped \\
anisotropic S=1 chain antiferromagnet NiCl$_{2}$-4SC(NH$_{2}$)$_{2}$}
\author{T.~A.~Soldatov}
\affiliation{P.~L.~Kapitza Institute for Physical Problems, RAS, 119334 Moscow, Russia} \affiliation{Moscow Institute for
Physics and Technology, 141701 Dolgoprudnyi, Russia}

\author{A.~I.~Smirnov}
\affiliation{P.~L.~Kapitza Institute for Physical Problems, RAS, 119334 Moscow, Russia} \affiliation{National Research
University Higher School of Economics, 101000 Moscow, Russia}

\author{K.~Yu.~Povarov}
\affiliation{Laboratory for Solid State Physics, ETH Z\"{u}rich, 8093 Z\"{u}rich, Switzerland}

\author{A.~Paduan-Filho}
\affiliation{High Magnetic Field Laboratory, University of S\~{a}o Paulo, BR-05315-970 S\~{a}o Paulo, Brazil}

\author{A.~Zheludev}
\affiliation{Laboratory for Solid State Physics, ETH Z\"{u}rich, 8093 Z\"{u}rich, Switzerland}

\begin{abstract}

%\vspace{1 cm}

   We studied electron spin resonance in a quantum magnet  NiCl$_{2}$-4SC(NH$_{2}$)$_{2}$, demonstrating a field-induced quantum phase transition
   from a quantum-disordered phase  to an antiferromagnet. We observe two branches of the antiferromagnetic resonance of the ordered phase, one of them
   has a gap and the other is a Goldstone mode with zero frequency at a magnetic field along the four-fold axis. This zero frequency mode
   acquires a  gap at a small tilting of the magnetic field with respect to this direction.
   %This angular dependence gives an alternative explanation
   %for  the low
   %frequency resonances, which earlier was ascribed to an exchange mode of a multisublattice structure.
   The upper gap was found to be
   reduced  in the  doped compound Ni(Cl$_{1-x}$Br$_{x}$)$_2$-4SC(NH$_{2}$)$_{2}$ with $x=0.21$. This reduction is
   unexpected because of the previously reported rise of the main exchange constant in a doped compound.
   Further, a nonresonant diamagnetic susceptibility $\chi^{\prime}$ was found for the ordered phase in a wide
   frequency range above the quasi-Goldstone mode. This dynamic diamagnetism is  as large as the dynamic
   susceptibility of the  paramagnetic resonance. We speculate that it originates from a two-magnon absorption band of low-frequency dispersive magnon
   branch.

\end{abstract}

\date{\today}

\maketitle

\section{Introduction}\label{Introduction}

Antiferromagnetic spin  $S = 1$ chains were intensively studied both theoretically and experimentally. Their peculiar features are
quantum-disordered nonmagnetic ground states with an energy gap for magnetic excitations.  The spin gap of collective excitations may have an
exchange origin, as the well known  Haldane gap~\cite{Haldane1}, or, alternatively, may be due to the single ion easy-plane anisotropy. The
anisotropy makes the single-ion states with $S^z=0$ preferable, preventing the  magnetic ordering~\cite{Sakai2,Wierschem1,Wierschem2}. While a lot
of Haldane magnets were studied, quantum paramagnet ground states provided by strong anisotropy are not typical. NiCl$_{2}$-4SC(NH$_{2}$)$_{2}$
(dichloro-tetrakisthiourea-nickel(II), abbreviated as DTN) is a rare example of a spin system of this type.
     The low-temperature quantum disordered state of  DTN was found  to be unstable in the magnetic field $H_{c1}= 21$~kOe directed parallel
     to the four-fold axis $c$.
     In this field
a quantum phase transition into the antiferromagnetically ordered phase occurs~\cite{Paduan-Filho1}. At this transformation the spin gap of
 excitations at the Brillouin zone boundary is closing~\cite{Tsyrulin}. In the field range $H_{c1} < H < H_{c2}$  DTN is an antiferromagnet
with magnon excitations. The antiferromagnetic state disappears upon heating above the N\'{e}el temperature $T_N$ which has a maximum value
$1.2$~K in the magnetic field near $70$~kOe. At zero temperature  the values of the critical fields are $H_{c1} = 21$~kOe and $H_{c2} = 126$~kOe~
\cite{Paduan-Filho1,Paduan-Filho2,Zapf,Paduan-Filho3}. The second critical field of the field-induced antiferromagnetic phase at $T=0$ coincides
with  the field of magnetic saturation.  At the temperature rise both critical fields  move  to the middle of the interval, i.e. to $70$~kOe
  and disappear above $1.2$~K. The field-induced ordering in DTN was interpreted, in particular,  in terms of Bose-Einstein condensation
  of magnons~\cite{Zapf},  although the validity of such an interpretation has been later questioned \cite{Wulf2015}.

 A quantum phase transition in DTN was recently shown to be induced not only by a magnetic field but also by non-magnetic Br-doping.
  Br-ions substitute
   Cl-ions and this creates bond disorder via local variations of exchange and anisotropy parameters.
  Zero-field neutron scattering experiments~\cite{Wulf,Povarov1,Povarov2,Mannig} reveal a  decrease of the  spin excitation gap
 at the boundary of  the Brillouin zone for %\\
 Ni(Cl$_{1-x}$Br$_{x}$)$_2$-4SC(NH$_{2}$)$_{2}$  (DTNX). This gap is completely closed at $x\simeq0.16$~\cite{Povarov1,Povarov2,Mannig}.
 At $x = 0.21$ the antiferromagnetic long range order  appears
 even at zero field at $T_{N} = 0.64$~K~\cite{Povarov2}.

The Bose-Einstein scenario and the easy plane antiferromagnetic ordering itself should result in a gapless Goldstone mode in a field ${\vec H
}\parallel c$~\cite{Zapf,Zvyagin2}. A low-frequency  electron spin resonance (ESR) mode was indeed observed in the ordered phase, in addition to a
higher resonance branch as described in Ref.~\cite{Zvyagin2}. However, because of the nonzero frequency, the low-frequency resonance mode was
interpreted as exchange mode with anti-phase oscillations of two interpenetrating tetragonal subsystems. At the same time, a Goldstone mode was
assumed to have a zero frequency and to remain invisible. The unusual low-frequency mode as well as the essential influence of nonmagnetic
impurities on the spin excitations spectra, found in neutron scattering experiments, motivated us to perform a detailed ESR investigation at a
wider frequency range, different orientations of the magnetic field and different doping concentrations. In the present work we find that the
previously observed low-frequency antiferromagnetic resonance mode appears at a finite frequency due to a small tilting of magnetic field with
respect to $c$-axis. This breaks the four-fold axis symmetry and, hence, results in the gap for the quasi-Goldstone mode. This mode drops to zero
frequency at the precise orientation. Thereby the two-branch antiferromagnetic resonance spectrum with a gapped mode and a true Goldstone mode is
experimentally confirmed. The upper gap of the antiferromagnetic resonance was found to be reduced in the Br-doped sample with $x=0.21$, despite
the rise of the exchange integral near Br-impurity~\cite{NMR} and the widening of the magnon band~\cite{Mannig}.

In addition, our measurements of the microwave magnetic responses  both in DTN and DTNX
 reveal an unusual non-resonant dynamic magnetic
susceptibility of  a diamagnetic type. This dynamic diamagnetism appeared in a wide frequency range with a lower boundary at a quasi-Goldstone
mode. This nonresonant response  may indicate a band of two-magnon microwave absorption. The width of this band corresponds to the dispersion
range of the lower magnon branch.

\section {Crystal structure and magnetic parameters of DTN} \label{Parameters}

The crystals of DTN belong to the I4 tetragonal space group~\cite{Lopez-Castro}. Lattice parameters are $a = 9.558 \ \mathring{\textnormal{A}}$
and $c = 8.981 \ \mathring{\textnormal{A}}$. Magnetic ions Ni$^{2+}$ carrying spin $S=1$ are placed on a body-centered tetragonal lattice
consisting of two interpenetrating tetragonal sublattices. The sublattices are shifted relative to one another along the spatial diagonal of the
tetragonal unit cell by  half of its diagonal length.

 Each sublattice represents a spin subsystem with antiferromagnetic exchange. For the whole spin system
 the model  Hamiltonian at the magnetic field parallel to $c$-axis can be written as follows

\begin{eqnarray}
\mathcal{H} &=& \frac{1}{2} \sum\limits_{{\vec m}, {\vec \delta}} J_{\vec \delta}{\vec S}_{{\vec m}}\cdot{\vec S}_{{\vec m}+{\vec \delta}} +
\sum\limits_{{\vec m}}[D(S_{{\vec m}}^{z})^2 + g_{c}\mu_{B}HS_{{\vec m}}^{z}]\nonumber \\
&+& \frac{1}{2} \sum\limits_{{\vec n}, {\vec \delta}} J_{\vec \delta}{\vec S}_{{\vec n}}\cdot{\vec S}_{{\vec n+ \vec \delta}} +
 \sum\limits_{{\vec n}}[D(S_{{\vec n}}^{z})^2 + g_{c}\mu_{B}HS_{{\vec n}}^{z}] \nonumber \\
&+& \mathcal{H}_{int}  \label{Hamiltonian}
\end{eqnarray}

where $\vec{S}_{\vec{m},\vec{n}}$ are spin-1 operators at sites $\vec{m}$, $\vec{n}$ belonging to the first and second subsystem correspondingly,
 the vectors $\vec{\delta}$ connect the sites $\vec{m}$, $\vec{n}$ to their nearest neighbors within a subsystem, $D$ is the easy-plane single-ion
anisotropy constant, $J_{\vec{\delta}}$ are the exchange constants, $g_{c}\mu_{B}HS_{{\bf n,m}}^{z}$ are Zeeman terms, $g_c$ is the corresponding
$g$-factor and $\mathcal{H}_{\rm int}$ describes additional interactions, including  the interaction between the subsystems.

 The interaction between the subsystems is characterized by the exchange integral between
the corner and center cell ions $J^{\prime\prime}$. The exchange and anisotropy parameters were derived from the analysis of excitation spectrum
obtained in neutron scattering experiments on fully polarized samples~\cite{Tsyrulin}: $D = 8.9$~K, $J_c = 2.05$~K, $J_{ab}= 0.156$~K,
$J^{\prime\prime} = 0.08$~K . These values agree with the earlier ESR study~\cite{Zvyagin1} which report all these parameters except for
$J^{\prime\prime}$. The earlier neutron scattering investigations~\cite{Zapf} give close values  $D = 8.12$~K, $J_c = 1.74$~K, $J_{ab}= 0.17$~K.
The intersubsystem exchange $J^{\prime\prime}$ is frustrated. The magnetic subsystems are thus decoupled at zero field in the mean-field
approximation.

\section{Experimental details}\label{ExpDetails}

\begin{figure}[t]
\begin{center}
\vspace{0.1cm}
\includegraphics[width=0.45\textwidth]{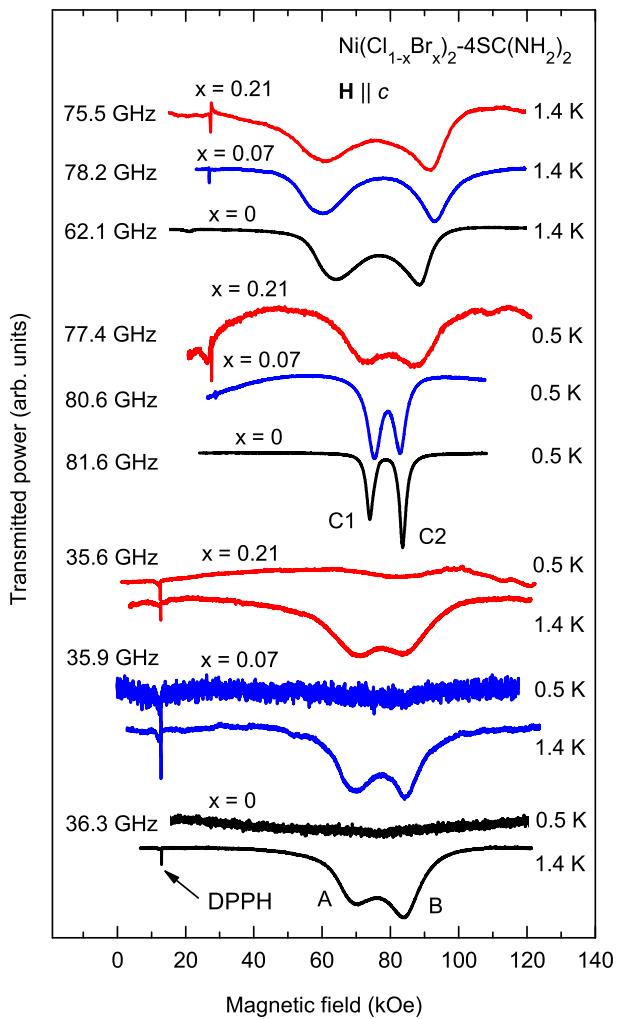}
 \caption{\label{fig2} (Color online) Examples of ESR lines obtained in DTNX with $x = 0, 0.07, 0.21$ at $T=0.5$~K and
 $T=1.4$~K  for ${\bf H }\parallel c$.}
\end{center}
\end{figure}

\begin{figure}[t]
\begin{center}
\vspace{0.1cm}
\includegraphics[width=0.46\textwidth]{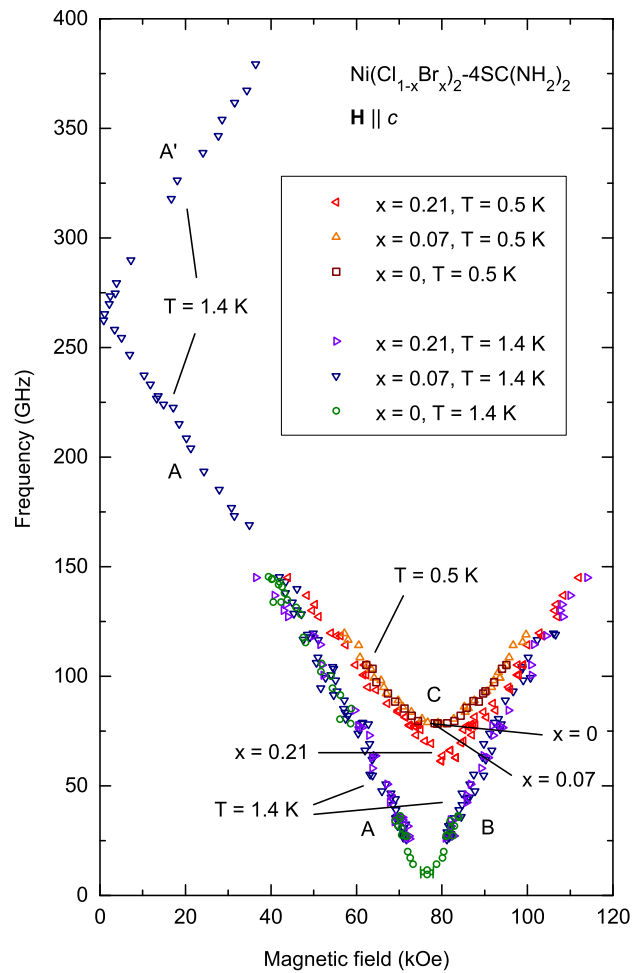}
 \caption{\label{fig1} (Color online) The frequency-field dependences of ESR in DTN and DTNX measured
 at  $T=0.5$~K and $T=1.4$~K for ${\bf H }\parallel c$. }
\end{center}
\end{figure}

\begin{figure}[t]
\begin{center}
\vspace{0.1cm}
\includegraphics[width=0.45\textwidth]{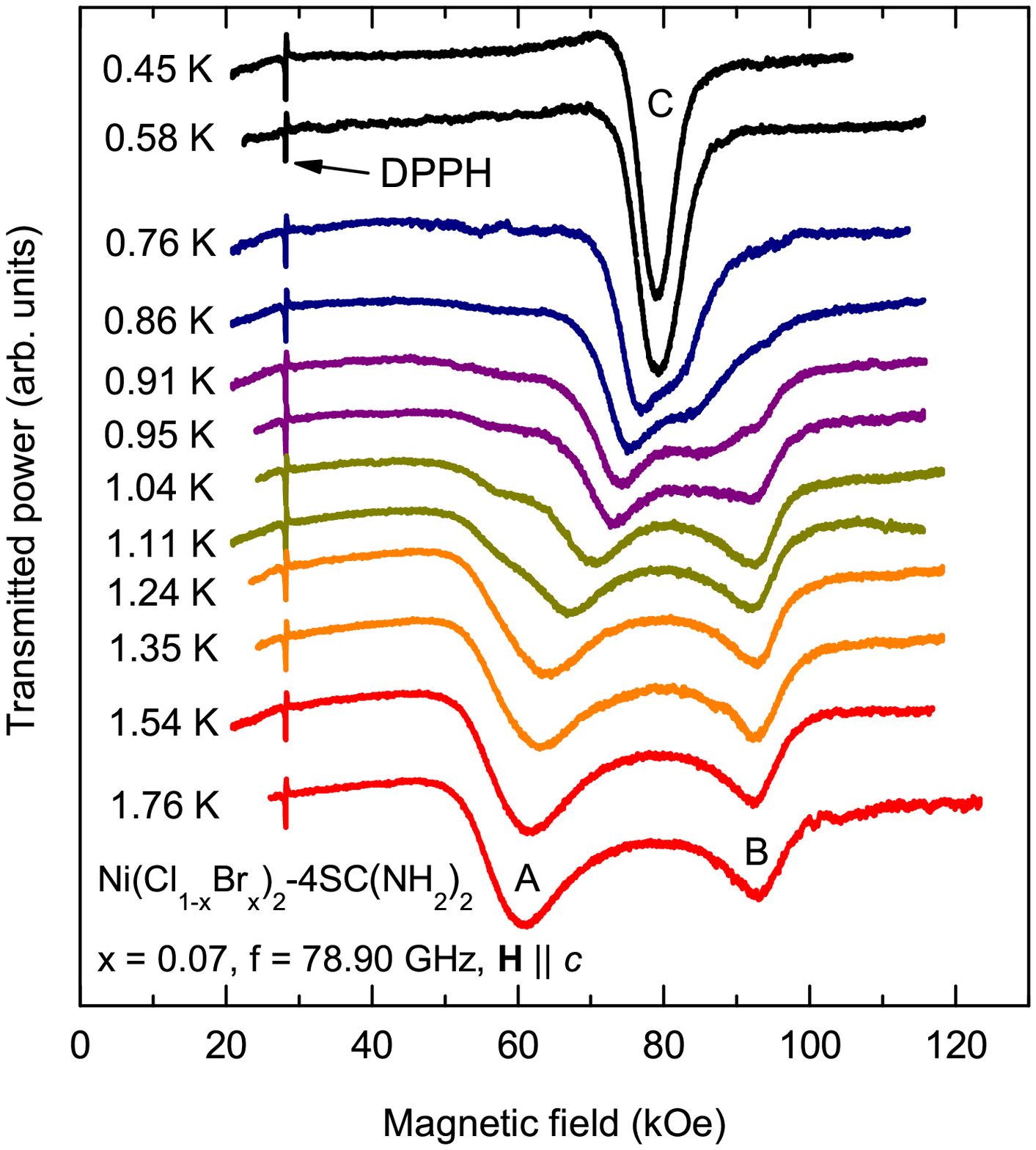}
 \caption{\label{fig3} (Color online) Temperature evolution of ESR spectra taken at $f = 78.90$~GHz and ${\bf H }\parallel c$
 in DTNX with $x = 0.07$.}
\end{center}
\end{figure}

\begin{figure}[t]
\begin{center}
\vspace{0.1cm}
\includegraphics[width=0.47\textwidth]{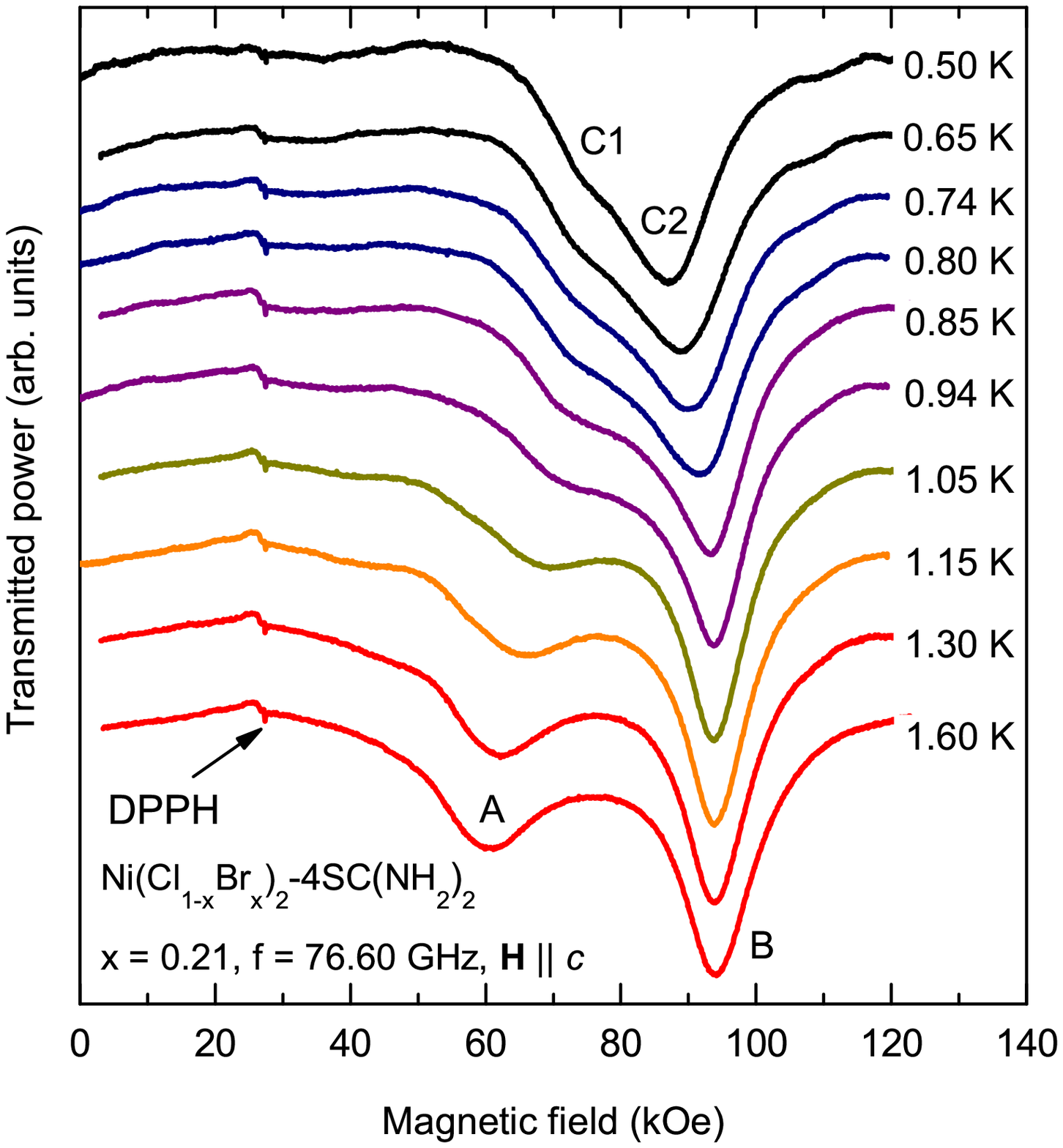}
 \caption{\label{fig4} (Color online) Temperature evolution of ESR spectra taken at $f = 76.60$~GHz and ${\bf H }\parallel c$
 in DTNX with $x = 0.21$. }
\end{center}
\end{figure}

\begin{figure}[t]
\begin{center}
\vspace{0.1cm}
\includegraphics[width=0.46\textwidth]{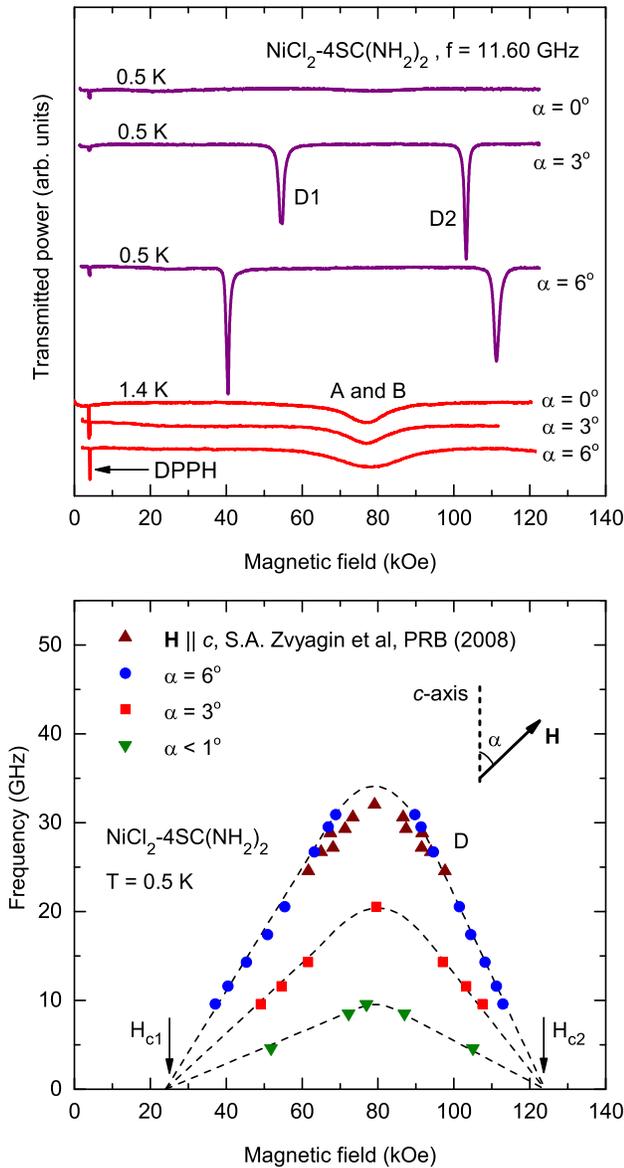}
 \caption{\label{fig5} (Color online) $11.6$~GHz ESR absorption curves at $0.5$~K and $1.4$~K (upper panel) and frequency-field
 diagrams for ESR in a low-frequency range at $0.5$~K (bottom panel) obtained in pure DTN for different orientations of
 the magnetic field in relation to $c$-axis.}
\end{center}
\end{figure}

\begin{figure}[t]
\begin{center}
\vspace{0.1cm}
\includegraphics[width=0.45\textwidth]{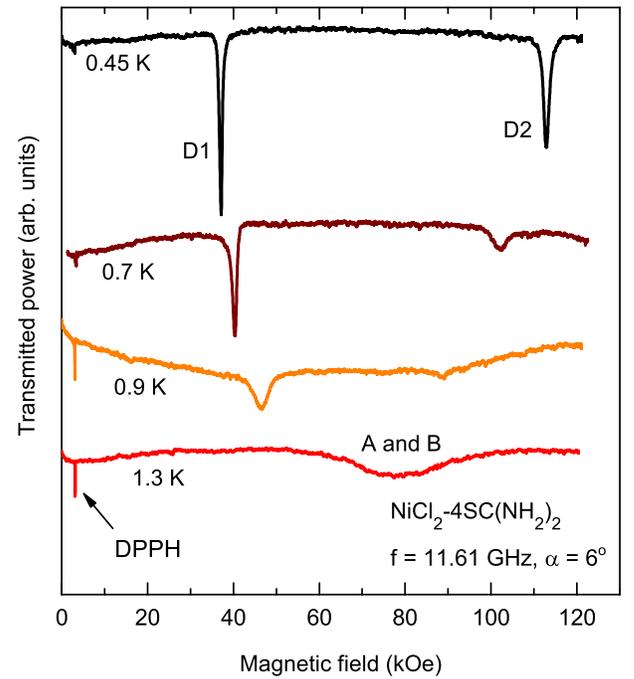}
 \caption{\label{fig6} (Color online) Temperature evolution of ESR line taken in pure DTN at frequency $f = 11.6$~GHz
 for the magnetic field tilted by 6 degrees from the $c$-axis.}
\end{center}
\end{figure}

\begin{figure}[h]
\begin{center}
\vspace{0.1cm}
\includegraphics[width=0.35\textwidth]{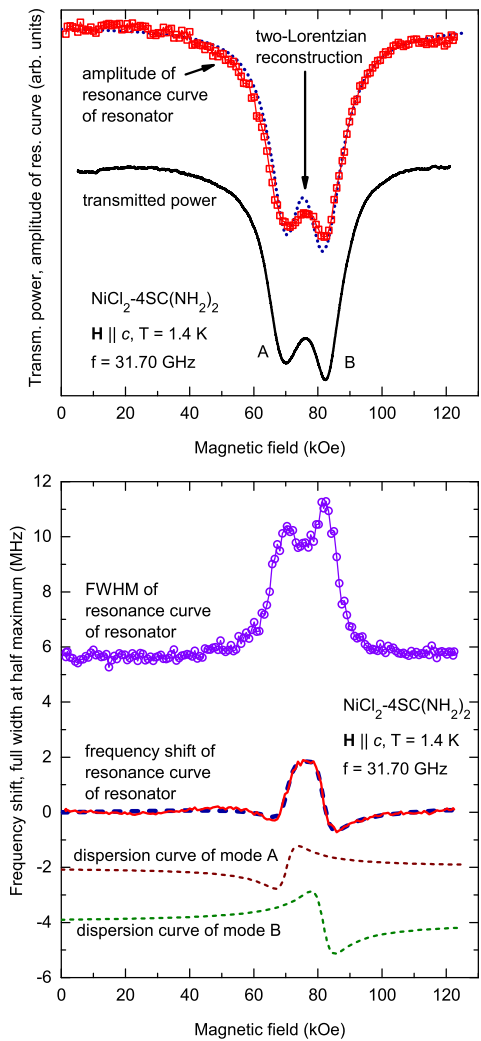}
 \caption{\label{fig9} (Color online) Upper panel: amplitude of resonator curve {\it vs} magnetic field
 for 31.70 GHz (symbols) and ESR-record at $T=$ 1.4 K (solid line). Dotted line is the two-Lorentzian model
 modified for large sample volume. Resonance fields and FWHM
 for  Lorentzians are the same as for dispersion curves on the lower panel.
 Bottom panel: Frequency shift and FWHM of the resonator {\it vs}  field.
Dashed line imposed on the frequency shift curve is a fit by two dispersion curves (dashed lines) for Lorentzian resonances. }
\end{center}
\end{figure}

\begin{figure*}[t]
\begin{center}
\vspace{0.1cm}
\includegraphics[width=0.95\textwidth]{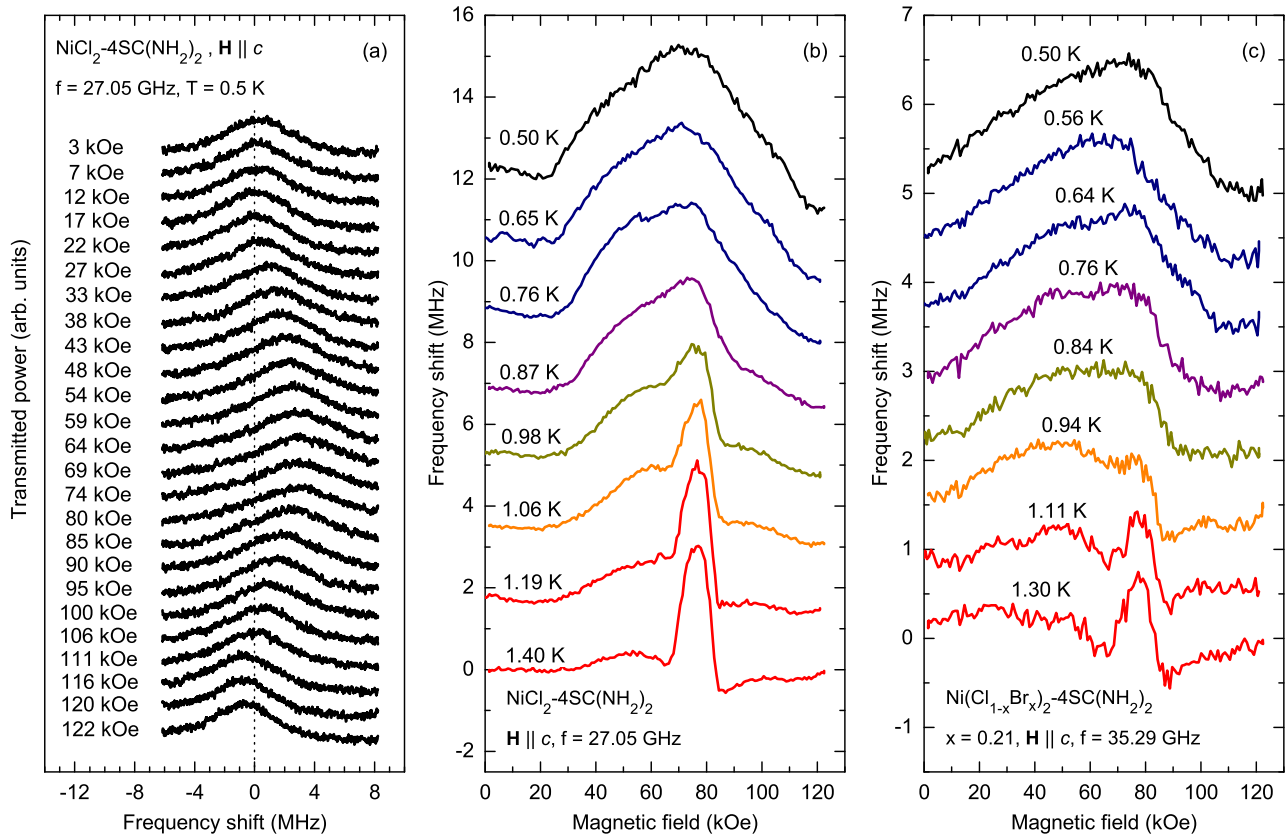}
 \caption{\label{fig7} (Color online) Panel (a): examples of resonance curves of resonator with DTN sample
 for $27.05$~GHz TE$_{112}$-mode at $0.5$~K and different values of field, ${\bf H }\parallel c$. Panel (b): temperature evolution
 of dependencies of resonance frequency shift on magnetic field, ${\bf H }\parallel c$, $f= 27.05$~GHz.
 Lines are shifted vertically by $1.75$~MHz to each other. Panel (c): temperature evolution
 of $\delta f_{res}$ for $35.29$~GHz (TM$_{021}$) mode of resonator with
 DTNX sample, $x = 0.21$. Lines are shifted vertically  by $0.75$~MHz to each other.}
\end{center}
\end{figure*}

\begin{figure*}[t]
\begin{center}
\vspace{0.1cm}
\includegraphics[width=0.95\textwidth]{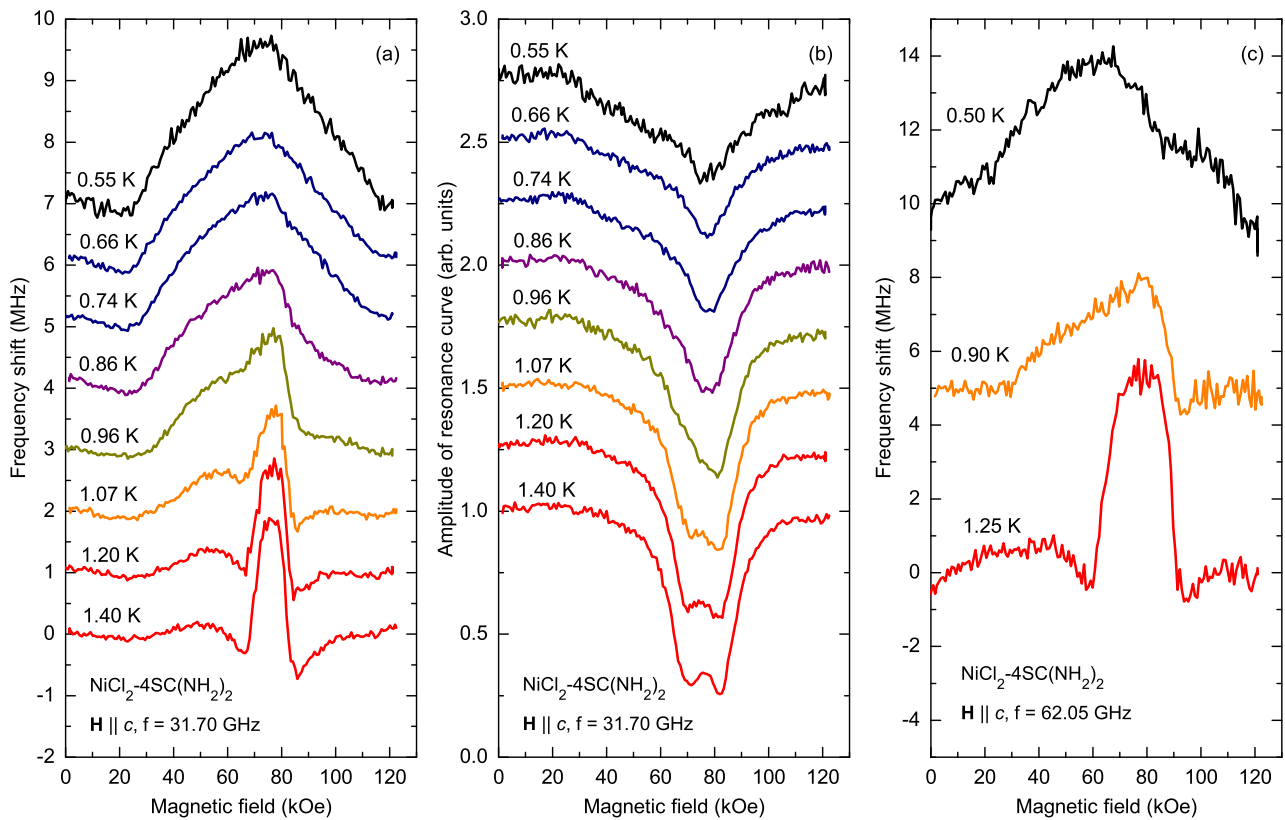}
 \caption{\label{fig8} (Color online) Panel (a): temperature evolution of the field-dependence  of the frequency shift $\delta f_{res}$ on magnetic
 field ${\bf H }\parallel c$ for $31.70$~GHz TM$_{211}$-mode. Lines are vertically shifted in turn by $1$~MHz. Panel (b): temperature evolution of
 the field-dependence of the amplitude of $31.7$~GHz mode. Lines are normalized to zero field amplitude for each of the temperature values
 and vertically shifted in turn by 0.25 a.u.  Panel (c): frequency shift for
  $62.05$~GHz mode of
 resonator. Lines are  vertically shifted in turn by $5$~MHz. }
\end{center}
\end{figure*}

\begin{figure}[t]
\begin{center}
\vspace{0.1cm}
\includegraphics[width=0.47\textwidth]{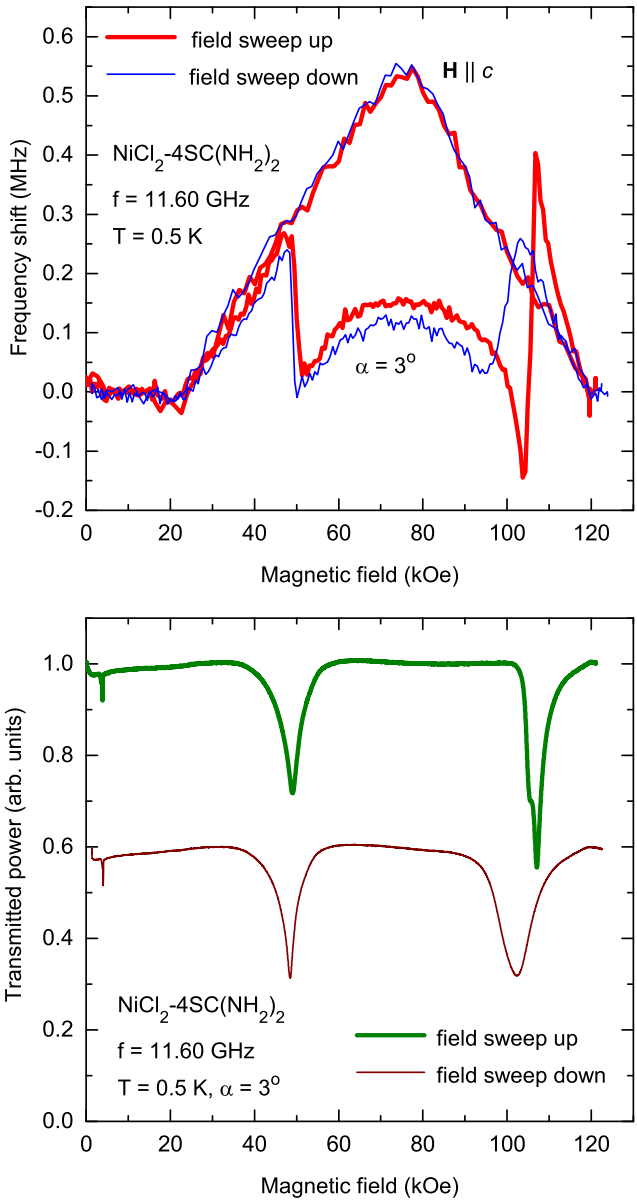}
 \caption{\label{fig10} (Color online) Dependencies of resonance frequency shift on magnetic field for $11.6$~GHz mode of
 resonator (upper panel) and  ESR absorption lines (bottom panel) measured  at $T=0.5$~K for ${\bf H }\parallel c$
 and for field tilted by 3 degrees from the $c$-axis. ESR lines are
  normalized to 1 and  the lower line is shifted down by 0.2}
\end{center}
\end{figure}

Single-crystalline samples of DTN and DTNX used in our experiments are from the same batches as studied in previous ESR and neutron scattering
experiments~\cite{Zvyagin1,Zvyagin2,Povarov1,Povarov2,Mannig}. Experimental results presented in Figs.~\ref{fig2}--\ref{fig4} and
Figs.~\ref{fig9}--\ref{fig10} were obtained using the crystals synthesized in ETH Z\"{u}rich and the results shown in Figs.~\ref{fig5},~\ref{fig6}
 -- for
crystals grown in University of S\~{a}o Paulo.

The ESR measurements were carried out with the use of a home-made transmission-type microwave spectrometer equipped with a superconducting
$120$~kOe magnet and $^{3}$He cryostat providing low temperatures down to $0.45$~K. Cylindrical, rectangular and cut-ring multimode resonators
were used to cover a wide frequency range $4 - 160$~GHz.  High-frequency $160 - 380$~GHz measurements were performed with use of a cylindrical
multimode resonator. Gunn diodes, back-wave oscillators and klystrons were used as microwave sources. Crystal samples were mounted inside the
resonators, and a small amount of 2,2-diphenyl-1-picrylhydrazyl (DPPH) was placed near the sample as a standard $g = 2.00$ marker for the magnetic
field. The sample size was chosen small enough not to disturb the electromagnetic field in a resonator. At the same time the sample should be
large enough to give the observable ESR signal. Thus, the crystals with the masses between 1 and 90 mg were used for measurements at different
frequencies.

\begin{figure}[b]
\begin{center}
\vspace{0.1cm}
\includegraphics[width=0.45\textwidth]{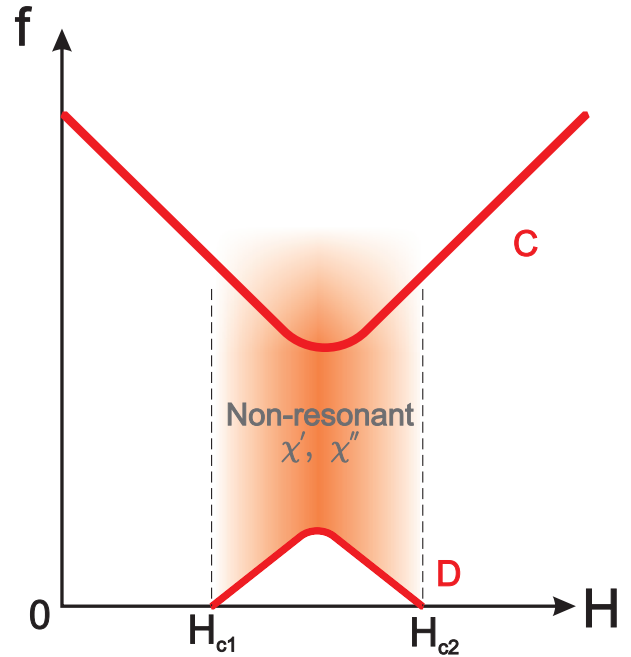}
 \caption{\label{Scheme} (Color online) Scheme of the frequency-field range of the nonresonant dynamic responce of DTN. }
\end{center}
\end{figure}

\begin{figure}[b]
\begin{center}
\vspace{0.1cm}
\includegraphics[width=0.45\textwidth]{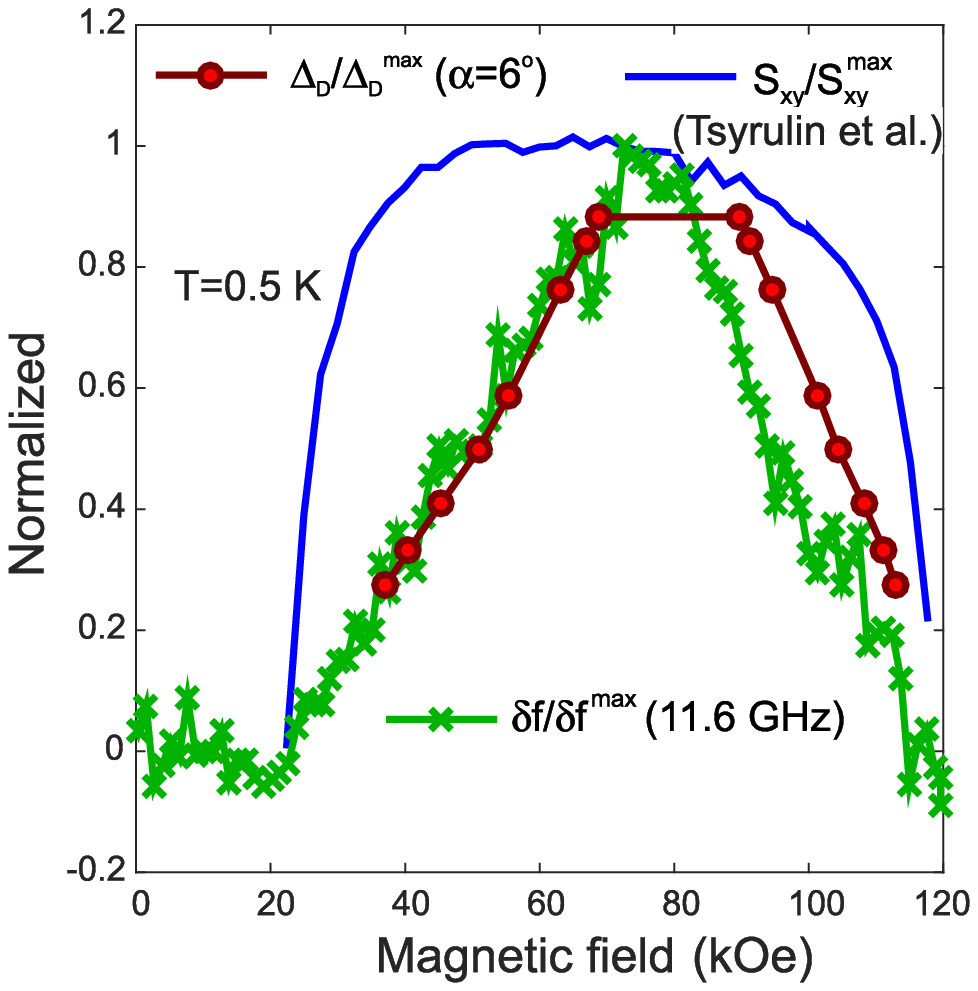}
 \caption{\label{Comparison} (Color online) Comparison of the field dependencies of the order parameter measured in
 neutron scattering experiments of Ref.~\cite{Tsyrulin} with the observed values of the resonator frequency shift
 and of the frequency of the quasi-Goldstone mode in a tilted field. }
\end{center}
\end{figure}

 ESR absorption lines were recorded as field-dependencies of the microwave power transmitted through the resonator.
If the generator is tuned to resonator frequency $f_{res}$, the transmitted signal is
\begin{eqnarray} U = U_{0}/(1+ 4 \pi \chi'' \eta Q)^2, \label{ImPartChi}
\end{eqnarray}
here $\eta = \int_{s} h^{2} dV / \int_{r} h^{2} dV$ is the filling factor, $h$ -- microwave magnetic field, the first and second integrals are
taken over the sample volume and over the resonator volume respectively,  $Q=f_{res}/(2 \delta_{\frac{1}{2}})$ is the quality factor of the empty
loaded resonator, $\delta_{\frac{1}{2}}$ is the half width at half maximium of the resonance curve,  $U_{0}$ is the signal passing through the
resonator in the absence of the sample. Typical value of $Q$ for the resonators is 3000-5000.

The change $\delta \chi '$ of the real part of the susceptibility with the magnetic field  may be derived from measurements of $f_{res}$ using the
following relation~\cite{Gurevich}
\begin{eqnarray}
\delta f_{res} = - 4 \pi \gamma \eta \delta \chi ' f_{0}, \label{RealPartChi}
\end{eqnarray}
here $\delta f_{res}$ is the  field-induced  shift of the $f_{res}$, $f_0$ is $f_{res}$ at zero field, $\gamma$
is a numerical coefficient which depends on the shape of the sample (for spherical sample it equals to 1). Using
this principles we derived the real and imaginary parts of the magnetic susceptibility of the sample  from
measurements of the resonance curve of the resonator.

\section{Experimental results}\label{ExpResults}

\subsection{ESR spectra at ${\bf H }\parallel c$}\label{FvsHdiagramsHc}

  The three   upper curves in Fig.~\ref{fig2} are taken at $T=1.4$~K and demonstrate examples of ESR absorption in the paramagnetic phase
 for samples with different dopant concentrations. The
 resonance fields presented in frequency-field diagram (Fig.~\ref{fig1})
 demonstrate that in the paramagnetic state
 the frequency-field dependence is the same for all dopant concentrations.
The observed paramagnetic spectrum has a nonzero frequency $\Delta_{1}$ in zero field, this mode is split by a magnetic field into two branches,
descending (A) and ascending (A$'$). The descending branch falls almost to zero and is replaced by a new ascending branch B. This spectrum is
similar to that of spin $S=1$ in an axial crystal field. For a spin in a crystal field the $f(H)$ dependencies of modes A, A$'$ and B are linear.
For DTN there is a weak deviation from linear dependence. The zero field frequency  we observe for $x=0.07$ sample is $265\pm 3$~GHz. This
coincides well with a value of $270 \pm 5$~GHz measured in ESR experiments of Ref.~\cite{Zvyagin2}and a gap of $1.06\pm0.04$~meV ($256 \pm
10$~GHz) measured in neutron experiments with pure samples \cite{Zapf}. The
 value of $\Delta_1$ for $x=0.21$, measured in neutron scattering
  experiments $1.1$~meV~\cite{Povarov2, Mannig} is almost the same.
We observe that branches A and B in the paramagnetic phase (at $T=1.4$~K) cross at a frequency of $10$~GHz.

 The temperature evolution of ESR lines  is presented in Fig.~\ref{fig3} and Fig.~\ref{fig4}. Upon cooling below $1.2$~K we observe a gradual
 convergence of
resonance modes A and B. At the low temperature $T=0.5$ K the two lines are closer, than in the paramagnetic phase, as presented in
Figs.~\ref{fig2},\ref{fig3},\ref{fig4}.  Near the middle of the field interval of the antiferromagnetic phase lines A and B merge, forming a mode
C. Mode C has a minimum frequency $\Delta_2$, which may be considered as AFMR gap.
%The measurements show that the gap stops to increase around 0.6 K and it is practically the same at 0.6 and 0.45 K.
For $x=0$ and $x=0.07$ samples the AFMR gap in the low temperature limit is $78$~GHz.   The doping with Br results in a broadening of AFMR line
and a change of the temperature induced shift of resonance lines. In a sample with $x=0.21$ the temperature-dependent shift of lines A and B is
smaller than in  $x=0$ and $x=0.07$ samples, the gap $\Delta_2$ is also smaller and equals 65~GHz instead of 78~GHz in a pure sample, see
Fig.~\ref{fig1}. For a strongly doped sample Fig.~\ref{fig4} demonstrates that 76~GHz ESR  is a doublet, corresponding to over-gap range. At the
same time the $x=0.07$ sample has a single ESR line indicating a gap frequency at 78 GHz, see Fig.~\ref{fig3}. This clearly proves a larger value
of the gap for the $x=0.07$ and $x=0$ samples. In addition, the minimum position for the gapped mode C shifts towards higher fields in the
$x=0.21$ sample for about $2$~kOe.
 In the sub-gap frequency range (below $\Delta_2$),  ESR signal completely disappears at cooling, see 36 GHz ESR lines
 in Fig.\ref{fig2}.

\subsection{ESR in a  tilted  field}\label{FvsHlowfreq}

In the ordered phase, at a precise orientation of the magnetic field along the $c$-axis, we do not observe any resonances at the frequencies below
the gap $\Delta_2$. Upon tilting the magnetic field with respect to $c$-axis, a pair of ESR lines appears. Their frequency increases with tilting.
ESR lines taken at different field angles relative to $c$-axis and the corresponding frequency-field dependencies are shown in Fig.~\ref{fig5}.
The larger the deviation of the field from $c$-axis   in $ac$-plane, the greater is the frequency of these modes (they are marked as modes D1,
D2). At changing the magnetic field, the frequency of D1,2-modes approaches zero at both boundary fields $H_{c1}, H_{c2}$ and reaches a maximum
value in the middle of the field range of the ordered phase, i.e. near $80$~kOe. We assume that these resonances have zero frequency at the exact
orientation ${\vec H }\parallel c$. We estimate the accuracy of sample orientation in our experiments as about 1$^\circ$. This may be due to the
disorientation of resonator with respect to the solenoid.

The temperature evolution of the pair of ESR lines in a magnetic field tilted by 6$^\circ$ away from $c$-axis is shown in Fig.~\ref{fig6}. It
demonstrates that the D-resonances exist only at low temperatures in the ordered phase.

\subsection{Dynamic diamagnetism in the ordered phase}\label{DynamDiam}

At cooling the samples below the  N\'{e}el temperature  we detect a significant change $\delta f_{res}$ of the resonance frequency of the
resonator in a range where ESR absorption is not observable. This low-temperature frequency shift  appears only in the field range of the ordered
phase, i.e. between the fields $H_{c1}$ and $H_{c2}$. We have measured $\delta f_{res}$ in different fields and for different resonance modes of
different resonators and found that the corresponding change of the real part of the dynamic magnetic susceptibility $\chi '$ appeared to be
negative (i.e. diamagnetic), and exists above the low-frequency mode D.

To characterize this observations quantitatively, we recorded resonance curves of the resonator in a field range from zero to $120$~kOe with a
step of about $0.5$~kOe. The analysis of resonance curves  gives dependencies of  resonator frequency,
 amplitude and width  on magnetic field. From these values we deduce changes of susceptibilities $\chi^\prime$ and $\chi^{\prime \prime}$.
 Capability  of this method
for the DTN samples was tested in the paramagnetic phase
 at ${\vec H }\parallel c$ and $T=1.4$~K for $31.70$~GHz  TM$_{211}$  mode of the cylindrical
resonator of a diameter 16 mm. The results are presented in Fig.~\ref{fig9}.

On the upper panel we see that the transmitted signal and the amplitude of the resonator resonance curve demonstrate the identical
field-dependencies. The frequency shift of the resonator is presented on the lower panel. This curve may be well fitted by two dispersion curves
of two Lorentzian resonances corresponding to modes A and B. The dispersion curves are known to represent the susceptibility $\chi^\prime$ for a
standard Lorentzian ESR curve. These dispersion curves, presented by dashed lines, are of inverted types because the A-frequency is  falling while
B-frequency is rising with magnetic field. The field-dependence of the amplitude of the resonator curve may be well reconstructed by two
Lorentzian curves with the same resonance fields and widths, as for dispersion curves. This reconstruction is presented  by a dotted line in the
upper panel of Fig.~\ref{fig9}. Thus, the observed response of the resonator to  changes of $\chi^\prime$ and $\chi^{\prime \prime}$ of the sample
corresponds well to the known changes of susceptibilities near paramagnetic resonance fields.

 The resonator curves for the ordered phase are presented
in panel (a) of Fig.~\ref{fig7}. These data are taken in different fields  ${\vec H }\parallel c$ and $T=0.5$~K for $27.05$~GHz TE$_{112}$ mode of
the same cylindrical resonator. One can see, that at low temperatures the increase of magnetic field causes a shift of the resonance curve towards
higher frequencies. This shift has a maximum at $H=80$~kOe and then comes back to zero at $H=120$~kOe. By a Lorentzian fit to these curves we
obtained the field dependence of $\delta f_{res}$. The positive value of $\delta f_{res}$ indicates a significant negative change of $\chi^\prime$
in the field interval between the critical fields $H_{c1}$ and $H_{c2}$. We estimated $\delta \chi '$ using Eq.~(\ref{RealPartChi}), the maximum
absolute value is about 0.2 emu/mol. From the comparison of low-temperature  $\delta f_{res}(H)$ curve (upper line in Fig.~\ref{fig7} (b)) and of
the  $\delta f_{res}$ curve of
 the paramagnetic phase (lower line), we conclude that the non-resonant low-temperature susceptibility  $\chi ^{\prime}$ has approximately  the
same value as one at paramagnetic resonance. Further, because $\chi ^{\prime}$ of the paramagnetic resonance exceeds the static susceptibility for
a $Q$-factor of the paramagnetic resonance\cite{Altshuler}(here $Q\simeq 2$), and considering the sum of contributions of two close lines A and B,
we conclude that $\chi^\prime \simeq$ 4$\chi_0\simeq$ 0.4 emu/mol. Here $\chi_0= 0.1 \ $emu/mol is the value of the order of low-temperature
differential static susceptibility \cite{Paduan-Filho1}. This is in a reasonable correspondence with the above estimate from the absolute value of
$\delta f_{res}$.

The amplitude of the resonance curve of the resonator demonstrates a reduction in the ordered phase, see Fig.~\ref{fig8}(b). This indicates the
increase of $\chi^{\prime \prime}$ of about 20\% of the maximum value of $|\chi^{\prime}|$. The dependence of the frequency shift of the resonator
on the magnetic field at $T=0.5$~K clearly shows that the shift appears between $25$  and $120$~kOe, marking the critical fields of the ordered
phase at $0.5$~K.  The temperature evolution of the dependencies of resonance frequency on magnetic field shown in panel (b) of Fig.~\ref{fig7}
demonstrates the decrease of $ \delta f_{res}$ and  $| \chi ' |$ at the temperature rise. At temperatures $T
> 1$~K the frequency shift $\delta f_{res}$ is observable only near the resonance fields of modes A and B of the paramagnetic phase.
A similar temperature evolution of the frequency shift dependence on magnetic field
 was measured in DTNX with $x = 0.21 \ $ at ${\vec H }\parallel c$
for $35.29$~GHz TE$_{012}$ mode (see panel (c) of Fig.~\ref{fig7}).
 These data additionally prove the coupling of dynamic diamagnetism to the order
parameter. They  illustrates that at $0.5$~K the peak position of the resonance curve starts to deviate exactly at zero field. This  confirms the
formation of the ordered state in zero field as manifested in Ref.~\cite{Povarov2}.
 We detected dynamic susceptibility at frequencies up to the upper AFMR branch C. An example of $\delta f _{res}$ records for the highest
  frequency of detailed measurements  $62.05$~GHz is given in Fig. \ref{fig8}, panel (c). Because of the uncertainty in the
 filling factor for different modes of the resonator when the electromagnetic wavelength is comparable to the size of the sample,  we are unable
 to  derive the  exact numerical value of  ${\chi^\prime}$ and its frequency dependence. Nevertheless, we claim it is always negative within the
 antiferromagnetic phase  and it is comparable with the dynamic susceptibility of the paramagnetic resonance at $T=1.4$~K within the range
 $9-75$~GHz.
 Measurement of the dynamic susceptibility at higher
frequencies is  prevented by a strong ESR response of branch C, as well as by the close mutual proximity of modes of resonator, which results in
their overlapping. For the frequencies of $99$ and $119$~GHz we observe  characteristic kinks in the transmitted power records at fields of $25$
and $120$~kOe which indicate start and finish of the nonresonant response still present at these frequencies.

To test the bottom of the frequency range of the dynamic diamagnetic  susceptibility, we have measured  $\delta f_{res}$
 at low frequencies using a right angle resonator of the size $40 \times 8 \times 17$~mm
  with TE$_{013}$, TE$_{012}$, TE$_{011}$  modes for the field tilted from the $c$-axis.
   As an example, the experimental results for measurements at the frequency $11.6$~GHz
 with the mode TE$_{012}$ are presented
 in the upper panel of Fig.~\ref{fig10}. For intermediate fields strictly
 between resonance positions of two type-D modes we observed the suppression of the effect of dynamic diamagnetism.
 Thus, when the frequency is lower than the frequency of the
lower AFMR mode D, the diamagnetic contribution to the dynamic susceptibility  practically disappears. From this observation at all three
frequencies we deduce that the negative  $\chi^{\prime}$ at low temperatures exists at frequencies down to that of the quasi-Goldstone mode, where
it abruptly vanishes.

We ascribe the observed nonresonant susceptibilities $\chi ^{\prime}$ and  $\chi ^{\prime \prime}$ to dynamic magnetization in the (ab)-plane,
because of the standard ESR configuration of microwave fields used, where the RF magnetic field is perpendicular to the static field. We measured
also the frequency shift of the resonator placing the sample at the maximum of the  microwave electric field, which is perpendicular to the
external field. Then the sample was positioned at the maximum of the longitudinal microwave field. In both cases the shift was not greater than
for the initial sample position at the maximum of the transverse microwave field.  At the same time the intensity of the paramagnetic ESR was
reduced only by a factor of 2 in comparison with the initial experiment.
   This indicates a presence of all components of the microwave fields
  in the sample, i.e. we have
  a mix of polarizations of microwave fields within the sample, and a quantitative polarization experiment is impossible because of a
   large size of the sample. The sample of a size of 3-4 mm is comparable
   to a quarter of the electromagnetic wave length in the dielectric sample with $\varepsilon \simeq $4. However, the
  fraction of the sample volume with a certain polarization of electric or magnetic microwave field was changed and no sufficient
  increase of the diamagnetic response was observed. This means that the observed nonresonant dynamic susceptibility is  mainly due to the transverse
  magnetization oscillation.

 \section{Discussion}

\subsection{Paramagnetic resonance and AFMR gap } \label{AFMRmodesHcDiss}

The paramagnetic resonance at  $T=1.4$~K is mainly analogous to the ESR spectrum of a spin $S=1$ in a crystal field. It has a gap in zero field
and a descending branch which reaches approximately zero frequency for a precise orientation of the magnetic field along the four-fold axis. We
observe the zero-field gap $\Delta_1=270$~GHz  and  a decreasing  branch with the minimum frequency of $10$~GHz. This nonzero frequency could be
ascribed to the misorientation of about 1$^\circ$,  our accuracy of the orientation of the sample.  Naturally,  this lower frequency might be also
affected by other weak interactions.

For the ordered phase we observe  a broadening of resonance lines under Br-doping of DTN which is naturally due to random variation of local
surrounding of magnetic ions (see Fig.~\ref{fig2}). The most significant modification of ESR spectrum caused by doping, is a decrease of the AFMR
gap $\Delta_2$.

In a molecular field theory of Ref.~\cite{Zvyagin2} the value of $\Delta_2$ is approximately proportional to $J$ and the field of the minimum of
the frequency of mode C  rises approximately linear in $D$. The calculation by method described in Ref.~\cite{Zvyagin2} results in the following
approximate dependence
   of  $\Delta_2$ and of the field of
   the minimum of AFMR frequency $H_{res}$ on $J$ and $D$:

   $\frac{\partial \Delta_2}{\partial J}$ = 45.7 GHz/K,   $\frac{\partial \Delta_2}{\partial D}$ = - 0.6 GHz/K

   $\frac{\partial H_{res}}{\partial J}$ = 6.9 kOe/K,     $\frac{\partial H_{res}}{\partial D}$ = 6.6 kOe/K

  These differential relations are valid in the vicinity of the experimental values $\Delta_2=78$~GHz, $H_{res}=8$~T.
From the observed decrease of $\Delta_2$  for $13$~GHz and an increase of $H_{res}$ for $2$~kOe we get the change of the exchange integral $
\delta J = - 0.3$~K and the change of anisotropy parameter of $ \delta D = 0.7$~K.
 These values
contradict to doping-induced changes of parameters $J$ and $D$  derived in Ref.~\cite{Mannig}. Here the zero-field dispersion curves of
excitations in DTNX where interpreted within a so-called generalized spin-wave theory~\cite{Matsumoto,Zhang}. This analysis of $x=0.21$ data
resulted in the opposite and much stronger changes: $\delta J=$0.75~K and $\delta D=-1.5$~K  with respect to the parameters of a pure compound
reported in Ref.~\cite{Tsyrulin}.

   Recently another approach for the calculation of the AFMR spectrum in the field-induced antiferromagnetically ordered
phase was suggested, which implies a $1/S$ expansion for the order parameter and resonance frequencies~\cite{ScherbakovUtiosov}. Here  a
significant influence of the interchain exchange is predicted while the effects of parameters  $J$ and $D$ on the gap $\Delta_2$ and $H_{res}$ are
of the same sign as in the theory of Ref.~\cite{Zvyagin2}.  In this approach, for DTN we have:

 $\frac{\partial \Delta_2}{\partial J} =$   125 GHz/K,     $\frac{\partial \Delta_2}{\partial D} =-$  40 GHz/K

   $\frac{\partial H_{res}}{\partial J} = $ 6.7 kOe/K,    $\frac{\partial H_{res}}{\partial D}$ = 3.3 kOe/K

This corresponds to  changes $ \delta J = - 0.1$~K and $ \delta D = 0.6$~K and is also in a contradiction with the above interpretation of
zero-field neutron scattering.

The above simple interpretation of the antiferromagnetic resonance in the doped samples appears to be controversial also because the effective
parameters of exchange and anisotropy, deduced from the observed AFMR frequency and field  correspond  to  a disordered phase on the
Sakai-Takahasi phase diagram~\cite{Sakai2,Wierschem2} (indeed, $D/J > $5 at $J_{ab}/J=$0.08) correspond to a quantum paramagnet instead of
observed antiferromagnetic phase of  this sample of DTNX. Both these contradictions indicate that AFMR in doped samples can not be described in
the same way as in the pure sample with just renormalised  parameters of $J$ and $D$.
  The approach to a strongly inhomogeneous system of DTNX as to a homogeneous one with renormalized parameters may be wrong, because
  the local variation of exchange and anisotropy at a Br-substituted site is more than 100\% as suggested by NMR study~\cite{NMR}. Local parameters for
  a magnetic ion near impurity are substantially affected and the influence of doping may be more complicated, than their renormalization.
An example of a cardinal change of the effective Hamiltonian by a nonmagnetic doping was given in Refs.~\cite{MaryasinZhitomirsky,Rb-KFePRL}. An
additional interaction in form of a biquadratic exchange was shown to appear in a frustrated system with a chaotic modulation of the exchange
network. In this case the effective Hamiltonian of the doped crystal should be corrected for additional terms.

\subsection{Quasi-Goldstone  AFMR  in a tilted field}\label{AFMRmodesHtiltedDiss}
 Low-frequency resonances observed in DTN in a  tilted  field clearly  originate from a zero frequency Goldstone mode, which should
exist at the exact orientation of magnetic field along $c$-axis~\cite{Zvyagin2}. At a tilting of the magnetic field the axial symmetry is lost and
the degeneracy of spin configurations with respect to  rotation around the $c$-axis is lifted. This should result in a nonzero frequency of spin
oscillations. Indeed, as may be seen in Fig. \ref{fig5}, with increasing the deviation angle the resonance frequency increases. The frequency of
this oscillations drops to zero at the boundary fields of the antiferromagnetic phase. This is natural since the order parameter also vanishes at
these fields. In contrast to the earlier conclusion of Ref.~\cite{Zvyagin2} we now believe, that the low-frequency resonance branch D appears due
to the tilting of the magnetic field and that is not an exchange branch of two interpenetrating antiferromagnetic systems.   The frequency of the
exchange mode in case of  frustration is an intriguing  and challenging problem. From a qualitative point of view the frequency of the exchange
mode is  probably much lower, since the corresponding exchange integral $J^{\prime\prime}$ is rather weak. Its value $0.08$~K corresponds to
$1.6$~GHz. Moreover, this interaction is frustrated, as the molecular fields of the corner-site spins cancel each other at the center-cell cite.
The expected exchange branch should have a frequency far below $1.6$~GHz, which is out of our frequency range and may be masked by weak
interactions.

\subsection{Dynamic diamagnetism in the ordered phase}\label{DynamDiamDiss}

The nature of the observed dynamic diamagnetism of the ordered phase of DTN is  not clear at the present moment. A simple example of dynamic
diamagnetism is given by a conventional ESR in a paramagnet, where  a negative $\chi^{\prime}$ exists in a narrow range above the ESR frequency,
see, e.g. Ref.~\cite {Altshuler}. This kind of diamagnetism should exist only near the resonance frequency and disappear below the frequency of
ESR. In our experiments we observe a strong diamagnetic response in a wide frequency range. This range has a sharp lower boundary at the frequency
of mode D. In the upper direction it spreads at least to the frequency of  about 100 GHz.
  The frequency-field range of the nonresonant dynamic
susceptibility is schematically shown in Fig. \ref{Scheme}. To explain the observation of the intensive dynamic diamagnetic susceptibility in a
wide frequency range, we propose this is due to a coupling of microwave field to magnon pairs.
 Two kinds of this coupling were suggested, see, e.g., Refs.~\cite{Suhl,Gurevich}.
  For the coupling of the first type the power of the microwave pumping of the frequency $f$ is absorbed by pairs of magnons with a half frequency
 $f/2$ and opposite wavevectors ${\vec k}$. This coupling results in the parametric excitation of magnons when the microwave field
 exceeds a threshold value. The parametric excitation is also known as Suhl's instability of the first kind.
 The parametric resonance condition $f=f({\vec k})+f({-\vec k})=2f({\vec k})$ is necessary for this coupling.  Above the threshold the magnon number rises
  exponentially in time and finally it is limited by an
additional nonlinearity at a more high level. Below the threshold, in a stationary regime, thermally activated pairs of  magnons absorb the power
from the pumping which results in the subthreshold absorption.
 This effect corresponds to a microwave susceptibility in a wide range between the doubled minimum and doubled maximum frequencies of the magnon branch.
 The theory of subthreshold absorption was suggested in Ref.~\cite{KaganovTsukernik}, experimental observations are reported in,
 e.g.,  Ref.~\cite{ProzorovaSmirnov,Yamazaki,Pereverzev}.

  Another coupling of  a microwave magnetic field and pairs of magnons is the reason for  Suhl's instability of
  the second kind. It results in the saturation of ferromagnetic resonance. Two quanta of microwave pumping are converted
  into two magnons with opposite wave vectors ${\vec k}$ and ${-\vec k}$  corresponding to a resonance condition $2f=2f({\vec k})$.
  This coupling \cite{Suhl} originates from
  a nonlinear interaction of the uniform spin precession excited by microwave field with the spin wave modes with wavevectors ${\vec k}, {-\vec k}$.
   The frequency range
  for the microwave absorption due to this mechanism  is strictly between the bottom and the top of spin-wave branch.
    The subthreshold absorption due to  coupling of  Suhl's second kind is, in principle, possible, but not confirmed
 in experiments.

  For  DTN
 we may suppose the upper  AFMR branch C as a source of the uniform spin oscillations, which are excited at a lower wing of its resonance curve
 (it has a linewidth of  about $15$~GHz). Then, pairs of magnons of D-branch may be excited by the coupling of the second kind.
 The magnons of D-branch have a dispersion  range which may be estimated as
 $2J_c \sqrt{S(S+1)} \simeq 110$~GHz. This estimate is in  correspondence with the observed\cite{Tsyrulin} dispersion range $0.4$~meV $= 97$~GHz
 of the lower magnon branch in a magnetic field of $60$~kOe.
  The range of two-magnon absorption with the coupling of second kind
  corresponds well with the range of the nonresonant susceptibility observed in DTN. We identify  the lower frequency
  limit exactly on
 the AFMR frequency of D-mode and see that the whole range propagates above $100$~GHz.
 At a frequency below the uniform precession of the mode D this coupling should be terminated, as we observe
 in the experiment with the tilted field,  when the dynamic diamagnetism drops at a field of D-type resonance.

 The subthreshold absorption observed in Refs.~\cite{ProzorovaSmirnov,Yamazaki,Pereverzev}
  was recorded as imaginary part of the dynamic susceptibility.  The peculiar features of wide-range dynamic susceptibility of DTN are
   that the coupling is of the  Suhl's second kind,
      diamagnetic susceptibility $\chi^\prime$ dominates over the
      imaginary part $\chi^{\prime\prime}$, and the low-frequency cut-off of the effect corresponds to the resonance frequency of the uniform mode
      and not to the doubled frequency of this mode.
      The imaginary part of dynamic susceptibility should be inevitably accompanied by the real part according to Kramers-Kronig relations. This
      justifies the real part of the dynamic susceptibility.
       The dynamic  susceptibility for the parametric excitation (Suhl's instability of the first kind) of spin waves in an antiferromagnet,
 was also found to have a  significant diamagnetic real part~\cite{ProzorovaKveder}.

It is worth to note that the diamagnetic susceptibility and the frequency of the quasi-Goldstone mode in a tilted field  have practically
coinciding field dependencies. Both $\chi^{\prime}$ and $f_D$ have zero values at the boundaries of the antiferromagnetic phase and demonstrate a
maximum in a form of an apex of a triangle. The order parameter also is zero at the critical fields $H_{c1,2}$. However, it has a wide maximum in
the middle of the range.  The comparison of the field-dependencies of these three parameters is shown in Fig. \ref{Comparison}.

 Summarizing the discussion of the
dynamic diamagnetism of DTN, we conclude, that we have only a quite hypothetical explanation for this effect. We  explain a wide-range dynamic
susceptibility via the coupling between the microwave pumping and pairs of magnons.  This coupling is probably of the same type as for the Suhl's
instability of the second type. In principle, other mechanisms of transformation of the uniform spin oscillation mode to spin waves, e.g. by
elastic scattering by defects might also result in a wide range of dynamic response with observed boundaries.

\subsection{Unsolved problems}\label{ProblemsDisc}

 We notice the following questions which are not yet discussed in theory, that would be of importance for
understanding of peculiar properties of DTN. First, this is the theory of the spin dynamics and AFMR of the quantum magnet beyond the critical
field of the field-induced ordering, including  the field range where the order parameter is weak. The second unsolved problem is the theory of
spin oscillations of an antiferromagnet with a body centered lattice with two interpenetrating antiferromagnetic subsystems, the exchange
interaction of which is frustrated in a molecular field approximation. It should, probably,  contain a consideration of order-by-disorder
mechanism for derivation of the ground state. The third question is the physical nature  of a continuum of  spin modes providing a wide range of
dynamic diamagnetism.

\section{Conclusions}

We have observed experimentally the following new features of the spin dynamics of chain $S=1$ anisotropic antiferromagnet
NiCl$_{2}$-4SC(NH$_{2}$)$_{2}$.

1. The spectrum of antiferromagnetic resonance consists of two branches, one of them is gapped and has the minimum frequency of $78$~GHz in the
low-temperature limit and the other is zero-frequency Goldstone mode.

2. The Goldstone mode acquires a gap at a small tilting of the field with respect to the four-fold axis of the crystal.

3. At a nonmagnetic doping, replacing 21 \% of Cl-ions by  Br-ions, the gap of the upper AFMR  branch is reduced.

4. There is a wide range of the strong dynamic diamagnetic response at the frequencies above the quasi-Goldstone mode. This dynamic diamagnetism
is proposed to appear due to a wide-range coupling of two-magnon states of a quasi-Goldstone mode D to  microwave magnetic field.

\section{Acknowledgements}
     We are indebted to A. K. Kolezhuk, A. V. Syromyatnikov and O. I. Utesov for valuable discussions and performing theoretical calculations
     of AFMR frequencies
     and giving us the results of their calculations before publication, S. A. Zvyagin for valuable comments, M. E. Zhitomirsky,
     L. E. Svistov, S. S. Sosin and V. N. Glazkov for numerous important discussions.
     The work at the Kapitza Institute (ESR experiments, microwave measurements and  data processing)
     is supported by the Russian Science Foundation, Grant No 17-12-01505.
   Construction and installation of the  $4.5$~GHz spectrometric unit is supported by the Russian Foundation for fundamental research
   (Grant No 19-02-00194). Analysis of two-magnon absorption is supported by the Program of Presidium of RAS.


\begin{thebibliography}{90}

\bibitem{Haldane1} F. D. M. Haldane, Continuum dynamics of the 1-D Heisenberg antiferromagnet: identification with the O(3)
nonlinear sigma model, Phys. Lett. B {\bf 93A}, 464 (1983).


\bibitem{Sakai2} T. Sakai, M. Takahashi, Effect of the Haldane
gap on quasi-one-dimensional systems, Phys. Rev. B {\bf 42}, 4537 (1990).

\bibitem{Wierschem1} K. Wierschem, P. Sengupta,
Quenching the Haldane Gap in spin-1 Heisenberg antiferromagnets, Phys. Rev. Lett. {\bf 112}, 247203 (2014).

\bibitem{Wierschem2} K. Wierschem, P. Sengupta, Characterizing the Haldane phase in quasi-one-dimensional spin-1 Heisenberg
antiferromagnets, Mod. Phys. Lett. B {\bf 28}, 1430017 (2014).

\bibitem{Paduan-Filho1} A. Paduan-Filho, X. Gratens, N. F. Oliveira, Field-induced magnetic ordering in
NiCl$_{2}$$\cdot$4SC(NH$_{2}$)$_{2}$, Phys. Rev. B {\bf 69}, 020405 (2004).

\bibitem{Tsyrulin}
N. Tsyrulin, C. D. Batista, V. S. Zapf, M. Jaime, B. R. Hansen, C. Niedermayer, K. C. Rule, K. Habicht, K. Prokes, K. Kiefer, E. Ressouche, A.
Paduan-Filho, M. Kenzelmann, Neutron study of the magnetism in NiCl$_{2}$-4SC(NH$_{2}$)$_{2}$, J. Phys.: Condens. Matter {\bf 25}, 216008 (2013).

\bibitem{Paduan-Filho3} A. Paduan-Filho, R. D. Chirico, K. O. Joung, R. L. Carlin, Field induced
magnetic ordering in uniaxial nickel systems: A second example, J. Chem. Phys. {\bf 74}, 4103 (1981).


\bibitem{Paduan-Filho2} A. Paduan-Filho, X.
Gratens, N. F. Oliveira, High-field magnetization in the quantum spin magnet NiCl$_{2}$4SC(NH$_{2}$)$_{2}$, J. Appl. Phys. {\bf 95}, 7537 (2004).

\bibitem{Zapf} V. S. Zapf, D. Zocco, B. R. Hansen, M. Jaime, N. Harrison,
C. D. Batista, M. Kenzelmann, C. Niedermayer, A. Lacerda, A. Paduan-Filho, Bose-Einstein Condensation of $S = 1$ Nickel Spin Degrees of Freedom in
NiCl$_{2}$-4SC(NH$_{2}$)$_{2}$, Phys. Rev. Lett. {\bf 96}, 077204 (2006).


\bibitem{Wulf2015} E. Wulf, D. Huvonen, R. Schonemann, H. Kuhne, T. Herrmannsdorfer, I. Glavatskyy, S. Gerischer, K. Kiefer, S. Gvasaliya,
A. Zheludev, Critical exponents and intrinsic broadening of the field-induced transition in NiCl$_2$-4SC(NH$_2$)$_2$,
 Phys. Rev. B {\bf 91}, 014406 (2015)


\bibitem{Wulf} E. Wulf, D. H\"{u}vonen, J.-W. Kim, A. Paduan-Filho, E. Ressouche, S. Gvasaliya, V. Zapf, A. Zheludev,
Criticality in a disordered quantum antiferromagnet studied by neutron diffraction, Phys. Rev. B {\bf 88}, 174418 (2013).

\bibitem{Povarov1} K. Yu. Povarov, E. Wulf, D. H\"{u}vonen, J. Ollivier, A. Paduan-Filho, A. Zheludev, Dynamics of a
bond-disordered S = 1 quantum magnet near z = 1 criticality, Phys. Rev. B {\bf 92}, 024429 (2015).

\bibitem{Povarov2} K.
Yu. Povarov, A. Mannig, G. Perren, J. S. M\"{o}ller, E. Wulf, J. Ollivier, A. Zheludev, Quantum criticality in a three
dimensional spin system at zero field and pressure, Phys. Rev. B {\bf 96}, 140414 (2017).

\bibitem{Mannig} A. Mannig, K. Yu. Povarov, J. Ollivier, A. Zheludev, Spin waves near the edge of halogen substitution induced magnetic order in
Ni(Cl$_{1-x}$Br$_{x}$)$_{2}$-4SC(NH$_{2}$)$_{2}$, Phys. Rev. B {\bf 98}, 214419 (2018).

\bibitem{Zvyagin2} S.
A. Zvyagin, J. Wosnitza, A. K. Kolezhuk, V. S. Zapf, M. Jaime, A. Paduan-Filho, V. N. Glazkov, S. S. Sosin, A. I. Smirnov, Spin dynamics of
NiCl$_{2}$-4SC(NH$_{2}$)$_{2}$ in the field-induced ordered phase, Phys. Rev. B {\bf 77}, 092413 (2008).

\bibitem{NMR} A. Orlova, R. Blinder, E. Kermarrec, M. Dupont, N. Laflorencie, S. Capponi, H. Mayaffre, C. Berthier, A. Paduan-Filho, M.
Horvati\'{c}, Nuclear Magnetic Resonance Reveals Disordered Level-Crossing Physics in the Bose-Glass Regime of the Br-Doped
Ni(Cl$_{1-x}$Br$_x$)$_2$-4SC(NH$_2$)$_2$ Compound at a High Magnetic Field, Phys. Rev. Lett {\bf 118}, 067203 (2017).

\bibitem{Lopez-Castro} A. Lopez-Castro, M. R. Truter, The
crystal and molecular structure of dichlorotetrakisthioureanickel, [(NH$_{2}$)$_{2}$CS]$_{4}$NiCl$_{2}$, J. Chem. Soc., 1309 (1963).

\bibitem{Zvyagin1} S. A. Zvyagin, J.Wosnitza, C. D. Batista, M. Tsukamoto, N. Kawashima, J. Krzystek, V. S.
Zapf, M. Jaime, N. F. Oliveira, A. Paduan-Filho, Magnetic Excitations in the Spin-1 Anisotropic Heisenberg Antiferromagnetic Chain System
NiCl$_{2}$-4SC(NH$_{2}$)$_{2}$, Phys. Rev. Lett. {\bf 98}, 047205 (2007).


\bibitem{Matsumoto} M. Matsumoto, M. Koga, Longitudinal spin-wave mode near quantum critical point due to uniaxial anisotropy, J. Phys. Soc. Jpn. {\bf 76}, 073709 (2007).

\bibitem{Zhang} Z. Zhang, K. Wierschem, I. Yap, Y. Kato, C. D. Batista, P. Sengupta, Phase diagram and magnetic excitations of anisotropic spin-one magnets, Phys.Rev.B {\bf 87}, 174405 (2013).

\bibitem{Gurevich} A. G. Gurevich, G. A. Melkov,  Magnetization Oscillations and Waves,
CRC Press 1996, section 5.3

\bibitem{ScherbakovUtiosov} A. S. Sherbakov,  O. I. Utesov,   to be published


\bibitem{MaryasinZhitomirsky} V. S. Maryasin and M. E. Zhitomirsky, Triangular Antiferromagnet with Nonmagnetic Impurities,
  Phys. Rev. Lett. {\bf 111}, 247201 (2013).

\bibitem{Rb-KFePRL}   A. I. Smirnov, T. A. Soldatov, O. A. Petrenko, A. Takata, T. Kida, M. Hagiwara, A. Ya. Shapiro, and M. E. Zhitomirsky
   Order by Quenched Disorder in the Model Triangular Antiferromagnet RbFe(MoO$_4$)$_2$, Phys. Rev. Lett. {\bf 119}, 047204 (2017).

\bibitem{Altshuler}  S. Al'tshuler, B. Kozyrev, Electron Paramagnetic Resonance, Academic Press (1964).

\bibitem{Suhl} H. Suhl, The Theory of Ferromagnetic Resonance at High Signal Powers, Phys. Chem. Solids, {\bf 1}, 209 (1957).

\bibitem{KaganovTsukernik} M. I.Kaganov, V. M. Tsukernik,  Nonresonance Absorption of Oscillating Magnetic Field Energy by a Ferromagnetic Dielectric, II
 JETP {\bf 11}, 952 (1960)

\bibitem{ProzorovaSmirnov} L. A. Prozorova , A. I. Smirnov
Microwave energy absorption by thermal magnons in the layered antiferromagnet BaMnF$_4$, JETP Letters {\bf 23},  130, (1976).

\bibitem{Yamazaki} H. Yamazaki, Parallel Pumping of the Brillouin-Zone-Boundary Magnons in  a  Two-Dimensional Ferromagnet K$_2$CuF$_4$.
Journal of the Physical Society of Japan, {\bf 37}, 667 (1974).

\bibitem{Pereverzev} Yu. V. Pereverzev, A. V. Stepanov, Two-Magnon Absorption in Low-Dimensional Biaxial Antiferromagnets.  FNT {\bf 3}, 502 (1977).

\bibitem{ProzorovaKveder} V. V. Kveder, L. A. Prozorova, Investigation of the Beyond-Threshold Susceptibility in Antiferromagnetic MnCO$_3$ and CsMnF$_3$
 in parametric excitation of spin waves,  JETP Letters {\bf 19}, 353, (1974).

\end{thebibliography}
\end{document}